\documentclass[aps,prd,preprint,superscriptaddress,tightenlines,%
nofootinbib]{revtex4}
\newcommand{\PRE}[1]{{#1}}   

\usepackage{amsmath}
\usepackage{amssymb}
\usepackage{color}
\usepackage{bm}
\usepackage{epsfig}
\usepackage{latexsym}

\newcommand{\postscript}[2]{\setlength{\epsfxsize}{#2\hsize}
   \centerline{\epsfbox{#1}}}
\newcommand{\comment}[1]{}

\def\MET{\mbox{${\hbox{$E$\kern-0.6em\lower-.1ex\hbox{/}}}_T$}}

\def\a{\alpha}
\def\thalf{\tfrac{1}{2}}

\def\al{\alpha}

\def\be{\beta}

\def\comment#1{{}}
\def\tfrac#1#2{{\textstyle\frac{#1}{#2}}}

\def\hat{\widehat}

\begin{document}

\preprint{
\hfil
\begin{minipage}[t]{3in}
\begin{flushright}
\vspace*{.4in}
MPP--2009--42\\
LMU--ASC 18/09\\
\end{flushright}
\end{minipage}
}

\vspace{1cm}
\title{LHC Phenomenology for String Hunters
\PRE{\vspace*{0.3in}} }

\author{Luis A. Anchordoqui}
\affiliation{Department of Physics,\\
University of Wisconsin-Milwaukee,
 Milwaukee, WI 53201, USA
\PRE{\vspace*{.1in}}
}

\author{Haim Goldberg}
\affiliation{Department of Physics,\\
Northeastern University, Boston, MA 02115, USA
\PRE{\vspace*{.1in}}
}

\author{Dieter L\"ust}
\affiliation{Max--Planck--Institut f\"ur Physik\\
 Werner--Heisenberg--Institut,
80805 M\"unchen, Germany
\PRE{\vspace*{.1in}}
}

\affiliation{Arnold Sommerfeld Center for Theoretical Physics\\
Ludwig-Maximilians-Universit\"at M\"unchen,
80333 M\"unchen, Germany
\PRE{\vspace{.1in}}
}

\author{Satoshi \nolinebreak Nawata}
\affiliation{Department of Physics,\\
University of Wisconsin-Milwaukee,
 Milwaukee, WI 53201, USA
\PRE{\vspace*{.1in}}
}
\affiliation{Max--Planck--Institut f\"ur Physik\\
 Werner--Heisenberg--Institut,
80805 M\"unchen, Germany
\PRE{\vspace*{.1in}}
}

\author{Stephan Stieberger}
\affiliation{Max--Planck--Institut f\"ur Physik\\
 Werner--Heisenberg--Institut,
80805 M\"unchen, Germany
\PRE{\vspace*{.1in}}
}

\author{Tomasz R. Taylor}
\affiliation{Department of Physics,\\
  Northeastern University, Boston, MA 02115, USA 
\PRE{\vspace*{.1in}}
}

\date{April 2009}
\PRE{\vspace*{.6in}}
\begin{abstract}\vskip 3mm
  We consider extensions of the standard model based on open strings
  ending on D-branes, with gauge bosons due to strings attached to
  stacks of D-branes and chiral matter due to strings stretching
  between intersecting D-branes.  Assuming that the fundamental string
  mass scale is in the TeV range and the theory is weakly coupled, we
  discuss possible signals of string physics at the Large Hadron
  Collider (LHC). In previous works, direct channel excitations of
  Regge recurrences in parton-parton scattering supplied the
  outstanding new signature.  The present work considers the deviation
  from standard model expectations for the 4-fermion processes
  $qq\rightarrow qq$ and $qq'\rightarrow qq'$, in which the
  $s$-channel excitation of string resonances is absent. In this case,
  we find that Kaluza-Klein recurrences at masses somewhat less than
  the string scale generate effective 4-fermion contact terms which
  can significantly enhance the dijet $R$ ratio above its QCD value of
  about 0.6. The simultaneous observation of a nearby resonant
  structure in the dijet mass spectrum would provide a ``smoking gun''
  for TeV scale string theory. In this work, we also show that $(1)$
  for $M_{\rm string}<3.5$~TeV, the rates for various topologies
  arising from the $pp\rightarrow Z^0 +$ jet channel  could
  deviate significantly from standard model predictions and $(2)$ that
  the sizeable cross sections for Regge recurrences can allow a
  $6\sigma$ discovery for string scales as large as 3 TeV after about
  1 year of LHC operation at $\sqrt{s}=10$~TeV and $\int {\cal L}\
  dt\sim 100$ pb$^{-1}$.
\end{abstract}

\thispagestyle{empty}
\setcounter{page}{0}
\maketitle

\section{Introduction}

Two of the most important scales in physics are the Planck scale
($M_{\rm Pl} = G_{\rm N}^{-1/2} \approx 10^{19}$~GeV) and the weak
scale ($M_{\rm W} \approx G_{\rm F}^{-1/2} \approx 300$~GeV), and there is a
longstanding problem in explaining the disparity between these scales.
The traditional view is to adopt $M_{\rm Pl}$ as {\em the} fundamental
scale and attempt to derive $M_{\rm W}$ through some dynamical
mechanism (e.g. renormalization group evolution).  In recent years,
however, a new framework with a diametrically opposite viewpoint has
been proposed, in which $M_{\rm W}$ is instead the fundamental scale
of nature~\cite{ArkaniHamed:1998rs}. D-brane string compactifications
with low string scale and large extra dimensions allow a definite
representation of this innovative premise~\cite{Antoniadis:1998ig}.

TeV-scale superstring theory provides a brane-world description of the
standard model (SM), which is localized on  membranes extending in $p+3$
spatial dimensions, the so-called D-branes. Gauge interactions emerge
as excitations of open strings with endpoints attached on the
D-branes, whereas gravitational interactions are described by closed
strings that can propagate in all nine spatial dimensions of string
theory (these comprise parallel dimensions extended along the
$(p+3)$-branes and transverse dimensions). The apparent weakness of gravity
at energies below few TeVs can then be understood as a consequence of
the gravitational force ``leaking'' into the transverse compact
dimensions of spacetime. This is possible only if the intrinsic scale
of string excitations is also of order a few TeVs. Should nature be so
cooperative, a whole tower of infinite string excitations will open up
at this low mass threshold, and new particles of spin $j$ follow the
well known Regge trajectories of vibrating strings: $j = j_0 + \alpha'
M^2$, where $\alpha'$ is the Regge slope parameter that determines the
fundamental string mass scale
\begin{equation}
M_s={1\over \sqrt{\alpha'}}\, .
\label{Ms}
\end{equation}

In a series of recent
publications~\cite{Anchordoqui:2007da,Anchordoqui:2008ac,
Anchordoqui:2008hi,Lust:2008qc,Anchordoqui:2008di}
we have computed open string scattering amplitudes in
D-brane models and have discussed\footnote{String Regge resonances in models 
with low string scale are also discussed 
in \cite{Cullen:2000ef,AAB}, while KK graviton exchange into the bulk, which
appears at the next order in perturbation theory, is discussed in
\cite{Cullen:2000ef,Dudas,ChialvaGT}.} the
associated phenomenological aspects of low mass string
Regge recurrences related to experimental searches for physics beyond
the SM at the LHC. In this work we extend our previous
discussion in three directions. The first is a calculation of
non-resonant $t$- and $u$-channel exchanges for the dominant $qq \to
qq$ subprocess contributing to hadronic dijets. The second is a
calculation of stringy contributions to $pp \to$ monojet, $pp \to$
dilepton + jet, and $pp \to $ trijet, which parallels our previous
analysis of $pp \to \gamma$ +
jet~\cite{Anchordoqui:2007da,Anchordoqui:2008ac}. The third is an
update of our dijet analysis to asses the range of discovery of
stringy resonances in the first run of the LHC at 10~TeV
center-of-mass energy.

We consider the extensions of the SM based on open strings ending on
D-branes, with gauge bosons due to strings attached to stacks of
D-branes and chiral matter due to strings stretching between
intersecting D-branes~\cite{Blumenhagen:2006ci}.  To develop our
program in the simplest way, we will work within the construct of a
minimal model in which we consider scattering processes which take
place on the (color) $U(3)_a$ stack of D-branes, which is
intersected by the (weak doublet) $U(2)_b$ stack of D-branes, as
well by a third (weak singlet) $U(1)_c$ stack of D-brane. These
three stacks of D(3+p)-branes entirely fill the uncompactified part
of space-time and wrap certain $p$-cycles $\Sigma^{(a,b,c)}$ inside
the compact six-dimensional manifold $M_6$. In the bosonic sector,
the open strings terminating on the $U(3)_a$ stack contain the
standard gluons $g$ and an additional $U(1)_a$ gauge boson $C$; on
the $U(2)_b$ stacks the open strings correspond to the weak gauge
bosons $W$, and again an additional $U(1)_b$ gauge field.  So the
associated gauge groups for these stacks are $SU(3)_{\rm C} \times
U(1)_a,$ $SU(2)_{\rm EW} \times U(1)_b$, and $U(1)_c$, respectively;
the physical hypercharge is a linear combination of $U(1)_a,$
$U(1)_b$, and $U(1)_c$, plus in general a forth $U(1)_d$, which is
not relevant for the following discussion.  The fermionic matter
consists of open strings, which stretch between different stacks of
D$(p+3)$-branes and are hence located at the intersection points.
Concretely, the left-handed quarks are sitting at the intersection
of the $a$ and the $b$ stacks, whereas the right-handed $u$ quarks
comes from the intersection of the $a$ and $c$ stacks and the
right-handed $d$ quarks are situated at the intersection of the $a$
stack with the $c'$ (orientifold mirror) stack. All the scattering
amplitudes between these SM particles, which we will need in the
following, essentially only depend on the local intersection
properties of these D-brane stacks.

Such intersecting D-brane models involve at least three
kinds of generic mass scales.  First, of course, there is the
fundamental string scale, given in (\ref{Ms}) in terms of the slope
parameter $\alpha'$. Second, compactification from ten to four dimensions
on an internal six--dimensional space of Volume $V_6$ defines a mass scale:
\begin{equation}
M_6={1\over V_6^{1/6}}\, .
\end{equation}
Third,  wrapping the color stack $a$ and the weak stack $b$ of
D$(p+3)$-branes (we are neglecting the other stacks for the moment)
around the internal $p$-cycles $\Sigma^{(a,b)}$ defines
two volumes $V_p^{(a,b)}$ of these sub-spaces and their associated
masses $(V_p^{(a,b)}=(2\pi)^p\ v_p^{(a,b)})$:
\begin{equation}
M_p^{(a,b)}={1\over (v_p^{(a,b)})^{1/p}}\, .
\end{equation}
In order to streamline our discussion of Kaluza-Klein and string winding
modes associated to internal $p$-cycles we make a simplifying assumption that
they are ``isotropic'' {\em i.e}.\ one needs only one scale to characterize
each $p$-cycle.

These three types of fundamental dimensional
parameters of D-brane models are linked to 4D physical observables in
the following way: First, the 4D Planck mass given by
\begin{equation}
M_{\rm Pl}^2=8\ e^{-2\phi_{10}}\ M_s^8\ \frac{V_6}{(2\pi)^{6}}
\end{equation}
determines the strength of gravitational interactions. Here, the
dilaton field $\phi_{10}$ is related to the string coupling constant
through $g_s=e^{\phi_{10}}$. Thus, for a string scale
$M_s\approx{\cal O}(1\,{\rm TeV})$, the volume of the internal space
$M_6$ needs to be as large as $V_6M_s^6={\cal O}(10^{32})$.  Second, the
4D gauge couplings of the strong and weak interactions are given in
terms of the volumes $V_p^{(a,b)}$ as
\begin{equation}
\label{Dpgaugecoupling}
g_{(a,b)}^{-2}=(2\pi)^{-1}\ M_s^p\ e^{-\phi_{10}}\ v_p^{(a,b)}\ .
\end{equation}
Again for a string scale $M_s\approx{\cal O}(1\,{\rm TeV})$ and using
the known values of the strong and weak gauge coupling constants:
$g_a^2/4\pi\equiv g_3^2/4\pi\approx 0.1$ and $g_b^2/4\pi\equiv
g_2^2/4\pi\approx {1\over 3}g_3^2/4\pi$ at the string scale
($g_2^2/4\pi=\alpha_{\rm em}/\sin^2\theta_W$,
$\sin^2\theta_W\approx0.23$, $\alpha_{\rm em}\approx 1/128$) we can
compute the volumes of the internal cycles, assuming
weak string coupling. To be specific, we choose $g_s=0.2$, and
then we obtain:
\begin{equation}
M_s^p\ v_p^{(a)}\approx 1\, ,\quad M_s^p\ v_p^{(b)}\approx 3\,.
\label{vol}
\end{equation}

The string mass scale $M_s$ and the volumes $V_p^{(a,b)}$ of the
internal cycles are closely related to the masses of those stringy
particles that can be possibly seen at the LHC.  First, there is the
infinite tower of open string Regge excitations of the known SM
particles.  These are completely model independent, i.e. independent
of the compactified dimensions.  They have the same
quantum numbers under the SM gauge group as the gluons and the quarks,
but in general higher spins, and their masses are just square-root-of-integer
multiples of the string mass $M_s$.  The first Regge excitations of
the gluon $(g)$ and quarks $(q)$ will be denoted by $g^*,\ q^*$,
respectively. The first excitation of the $C$ (the extra $U(1)$ boson
tied to the color stack) will be denoted by $C^*$. Only one assumption
will be necessary in order to set up a solid framework: the string
coupling must be small for the validity of perturbation theory in the
computations of scattering amplitudes. In this case, black hole
production and other strong gravity effects occur at energies above
the string scale, therefore at least the few lowest Regge recurrences are
available for examination, free from interference with some complex
quantum gravitational phenomena.  In fact, as discussed in
Refs.~\cite{Anchordoqui:2007da,Anchordoqui:2008ac,
Anchordoqui:2008hi,Lust:2008qc,Anchordoqui:2008di},
some basic properties of Regge resonances like their production rates
and decay widths are completely model-independent.

Second, in any D-brane compactification there is an infinite tower of
internal open string KK excitations of the SM fields along the
D$(p+3)$-branes. These particles have SM quantum numbers and, if the
internal cycles are isotropic, their masses can be written as
\begin{equation}
M_{KK}^{(a,b)}={n\over \bigl(v_p^{(a,b)}\bigr)^{1/p}}\ ,
\label{mkkvol}
\end{equation}
where the integer $n$ labels individual KK mass levels. Note
that these levels may be degenerate:  for instance, the lowest KK level is
typically $p$-fold degenerate in orientifold compactifications.
Using the same value for the string coupling constant we can, using Eqs.(\ref{vol})
and (\ref{mkkvol}), compute
the masses of the first ($n=1$) strong and weak KK excitations for two
possible brane scenarios, namely IIA orientifold with intersecting
D6-branes ($p=3$) and IIB orientifolds with wrapped D7-branes ($p=4$):
\begin{eqnarray}
&~&p=3:\quad M_{KK}^{(a)}\equiv M_{KK}^{(3)}\approx M_s\, ,
\quad M_{KK}^{(b)}\equiv M_{KK}^{(2)}\approx 0.70~M_s\, ,
\nonumber\\
&~&p=4:\quad  M_{KK}^{(a)}\equiv M_{KK}^{(3)}\approx M_s\, ,\quad M_{KK}^{(b)}
\equiv M_{KK}^{(2)}\approx 0.75~M_s\, .
\end{eqnarray}
The presence of these KK modes is a generic property of D-brane models
however their masses and their couplings to the SM particles depend
on the details of the compactification manifold.
Depending on the internal topology, there is
also the  possibility  of charged winding states around internal
1-cycles inside $\Sigma_{a,b,c}$.  If they exist, e.g. in
toroidal-like compactifications, their masses are proportional to the
radii of the wrapped cycles:
\begin{equation}
M_{\rm wind.}^{(a,b)}=m\ \bigl(v_p^{(a,b)}\bigr)^{1/p} \, M_s^2 \, .
\end{equation}
Typically the parameter $m$ is an integer and accounts for winding
numbers. More generally, $m$ may also be a positive real number comprising
further details of the D--brane setup like intersection angles or displacement.
Here again, one expects a certain degeneracy in each level.

Finally, in addition to the (universal) Regge, KK and winding modes, which are charged
under the SM, there typically exist several other neutral particles,
like transverse KK modes, several moduli scalars and other closed string
states. However, these states interact only gravitationally,
{\em i.e.} at string one--loop,
with the SM fields, and hence are not relevant for LHC physics.

The layout of the paper is as follows. Section~II presents an
overview of all formulae relevant for the $s$-, $t$-, and $u$-channel
string amplitudes of lowest massive Regge and Kaluza-Klein (KK)
excitations leading to dijets. Armed with the full-fledged string
amplitudes of all partonic subprocesses, in Sec.~III we study the main
features of the dijet angular distributions and we elucidate how
future measurements of these distributions can provide a potent method
for distinguishing between different compactification scenarios. In
Sec.~IV we study the various final-state topologies arising in the $pp
\to Z^0+$ jet channel. We show that, for $M_s <4$~TeV, monojet,
trijet, and dilepton plus jet configurations could provide
corroborative evidence for Regge excitations in D-brane models. In
Sec.~V we quantify signal and background rates of Regge recurrences in
the early LHC run, operating at 5~TeV per beam. We show that dijet
signals of string excitations at the LHC may be so large that they
could be exposed to in the early days of running. We summarize our
conclusions in Sec.~VI.

\section{Dijet scattering amplitudes}

\subsection{Regge recurrences}

The physical processes underlying dijet production in $pp$ and $p \bar
p$ collisions are the scattering of two partons $ij$, producing two
final partons $kl$ that fragment into hadronic jets. The corresponding
$2\to 2$ scattering amplitudes ${\cal M}(ij \to kl)$, computed at the
leading order in string perturbation theory, are collected in
Ref.~\cite{Lust:2008qc}. The amplitudes involving four gluons as well
as those with two gluons plus two quarks do not depend on the
compactification details and are completely model-independent.
All string effects are encapsulated in these amplitudes in one
``form factor'' function of Mandelstam variables $s,~t,~u$ (constrained
by $s+t+u=0$):
\begin{equation}
V(  s,   t,   u)= \frac{s\,u}{tM_s^2}B(-s/M_s^2,-u/M_s^2)={\Gamma(1-   s/M_s^2)\ \Gamma(1-   u/M_s^2)\over
    \Gamma(1+   t/M_s^2)}.\label{formf}
\end{equation}
The physical content of the form factor becomes clear after using the
well-known expansion in terms of $s$-channel resonances
\cite{gabriele}:
\begin{equation}
B(-s/M_s^2,-u/M_s^2)=-\sum_{n=0}^{\infty}\frac{M_s^{2-2n}}{n!}\frac{1}{s-nM_s^2}
\Bigg[\prod_{J=1}^n(u+M^2_sJ)\Bigg],\label{bexp}
\end{equation}
which exhibits $s$-channel poles associated to the propagation of
virtual Regge excitations with masses $\sqrt{n}M_s$. Thus near the
$n$th level pole $(s\to nM^2_s)$:
\begin{equation}\qquad
V(  s,   t,   u)\approx \frac{1}{s-nM^2_s}\times\frac{M_s^{2-2n}}{(n-1)!}\prod_{J=0}^{n-1}(u+M^2_sJ)\ .
\label{nthpole}
\end{equation}
In specific amplitudes, the residues combine with the remaining
kinematic factors, reflecting the spin content of particles exchanged
in the $s$-channel, ranging from $J=0$ to $J=n+1$.

In the following we isolate the contribution to the partonic cross
section from the first resonant state. The $s$-channel pole terms of
the average square amplitudes contributing to dijet production at the
LHC can be obtained from the general formulae given in
Ref.~\cite{Lust:2008qc}, using Eq.(\ref{nthpole}). However, for
phenomenological purposes, the poles need to be softened to a
Breit-Wigner form by obtaining and utilizing the correct {\em total}
widths of the resonances~\cite{Anchordoqui:2008hi}. After this is
done, the contributions of the various channels are as
follows~\cite{Anchordoqui:2008di}:
\begin{eqnarray}
|{\cal M} (gg \to gg)| ^2 & = & \frac{19}{12} \
\frac{g^4}{M_s^4} \left\{ W_{g^*}^{gg \to gg} \, \left[\frac{M_s^8}{(  s-M_s^2)^2
+ (\Gamma_{g^*}^{J=0}\ M_s)^2} \right. \right.
\left. +\frac{  t^4+   u^4}{(  s-M_s^2)^2 + (\Gamma_{g^*}^{J=2}\ M_s)^2}\right] \nonumber \\
   & + &
W_{C^*}^{gg \to gg} \, \left. \left[\frac{M_s^8}{(  s-M_s^2)^2 + (\Gamma_{C^*}^{J=0}\ M_s)^2} \right.
\left. +\frac{  t^4+  u^4}{(  s-M_s^2)^2 + (\Gamma_{C^*}^{J=2}\ M_s)^2}\right] \right\},
\label{gggg2}
\end{eqnarray}
\begin{eqnarray}
|{\cal M} (gg \to q \bar q)|^2 & = & \frac{7}{24} \frac{g^4}{M_s^4}\ N_f\
\left [W_{g^*}^{gg \to q \bar q}\, \frac{  u   t(   u^2+   t^2)}{(  s-M_s^2)^2 + (\Gamma_{g^*}^{J=2}\ M_s)^2} \right. \nonumber \\
 & + &  \left. W_{C^*}^{gg \to q \bar q}\, \frac{  u   t (   u^2+   t^2)}{(  s-M_s^2)^2 +
(\Gamma_{C^*}^{J=2}\ M_s)^2} \right]
\end{eqnarray}
\begin{eqnarray}
|{\cal M} (q \bar q \to gg)|^2  & = &  \frac{56}{27} \frac{g^4}{M_s^4}\
\left[ W_{g^*}^{q\bar q \to gg} \,  \frac{  u   t(   u^2+   t^2)}{(  s-M_s^2)^2 + (\Gamma_{g^*}^{J=2}\ M_s)^2} \right. \nonumber \\
 & + & \left.  W_{C^*}^{q\bar q \to gg} \, \frac{  u   t(   u^2+   t^2)}{(  s-M_s^2)^2 + (\Gamma_{C^*}^{J=2}\ M_s)^2} \right] \,\,,
\end{eqnarray}
\begin{equation}
|{\cal M}(qg \to qg)|^2  =  - \frac{4}{9} \frac{g^4}{M_s^2}\
\left[ \frac{M_s^4   u}{(  s-M_s^2)^2 + (\Gamma_{q^*}^{J=1/2}\ M_s)^2} + \frac{u^3}{(s-M_s^2)^2 + (\Gamma_{q^*}^{J=3/2}\ M_s)^2}\right],
\label{qgqg2}
\end{equation}
where $g$ is the QCD coupling constant $(\alpha_{\rm QCD}=\frac{g^2}{4\pi}\approx 0.1)$
 and $\Gamma_{g^*}^{J=0} = 75\, (M_s/{\rm TeV})~{\rm GeV}$,
$\Gamma_{C^*}^{J=0} = 150 \, (M_s/{\rm TeV})~{\rm GeV}$,
$\Gamma_{g^*}^{J=2} = 45 \, (M_s/{\rm TeV})~{\rm GeV}$,
$\Gamma_{C^*}^{J=2} = 75 \, (M_s/{\rm TeV})~{\rm GeV}$,
$\Gamma_{q^*}^{J=1/2} = \Gamma_{q^*}^{J=3/2} = 37\, (M_s/{\rm
  TeV})~{\rm GeV}$ are the total decay widths for intermediate states
$g^*$, $C^*$, and $q^*$ (with angular momentum
$J$)~\cite{Anchordoqui:2008hi}. The associated weights of these
intermediate states are given in terms of the probabilities for the
various entrance and exit channels
\begin{equation}
W_{g^*}^{gg \to gg} = \frac{8(\Gamma_{g^* \to gg})^2}{8(\Gamma_{g^* \to gg})^2 +
(\Gamma_{C^* \to gg})^2} = 0.44 \,,
\label{w1}
\end{equation}
\begin{equation}
W_{C^*}^{gg \to gg} = \frac{(\Gamma_{C^*
  \to gg})^2}{8(\Gamma_{g^* \to gg})^2 + (\Gamma_{C^* \to gg})^2} =
0.56 \, ,
\label{w2}
\end{equation}
\begin{equation}
W_{g^*}^{gg \to q \bar q}  = W_{g^*}^{q \bar q \to gg} =
\frac{8\,\Gamma_{g^* \to gg} \,
\Gamma_{g^* \to q \bar q}} {8\,\Gamma_{g^* \to gg} \,
\Gamma_{g^* \to q \bar q} + \Gamma_{C^* \to gg} \,
\Gamma_{C^* \to q \bar q}} = 0.71 \, ,
\label{w3}
\end{equation}
\begin{equation}
W_{C^*}^{gg \to q \bar q} = W_{C^*}^{q \bar q \to gg}  =
\frac{\Gamma_{C^* \to gg} \,
\Gamma_{C^* \to q \bar q}}{8\,\Gamma_{g^* \to gg} \,
\Gamma_{g^* \to q \bar q} + \Gamma_{C^* \to gg} \,
\Gamma_{C^* \to q \bar q}} = 0.29 \, .
\label{w4}
\end{equation}
Superscripts $J=2$ are understood to be inserted on all the $\Gamma$'s in
Eqs.(\ref{w1}), (\ref{w2}), (\ref{w3}), (\ref{w4}).
Equation~(\ref{gggg2}) reflects the fact that weights for $J=0$ and
$J=2$ are the same~\cite{Anchordoqui:2008hi}. In what follows we set
the number of flavors $N_f =6$.

\subsection{KK and winding modes}

The amplitudes for the four-fermion processes like quark-antiquark
scattering are more complicated because the respective form factors
describe not only the exchanges of Regge states but also of heavy
Kaluza-Klein and winding states with a model-dependent spectrum
determined by the geometry of extra dimensions~\cite{Chemtob:2008cb}.
Consider e.g. the
quark-quark scattering process. The square amplitude for the identical
quark flavors $qq\rightarrow qq$ is given by~\cite{Lust:2008qc}
\begin{eqnarray}
|{\cal M}(qq\rightarrow qq)|^2 &=& g^4 \left\{
{2\over 9}{1\over t^2}\Big[\big(sF^{bb}_{tu}\big)^2  +\big(sF^{cc}_{tu}\big)^2 +\big(uG'^{bc}_{tu}\big)^2  +
\big(uG'^{cb}_{tu}\big)^2     \Big] \right.
\nonumber \\ &+&
{2\over 9}{1\over u^2}\Big[\big(sF^{bb}_{ut}\big)^2  +\big(sF^{cc}_{ut}\big)^2 +\big(tG'^{bc}_{ut}\big)^2  +\big(tG'^{cb}_{ut}\big)^2\Big]\nonumber \\
 &-& \left. {4\over 27} {s^2\over tu}\big(
F^{bb}_{tu} F^{bb}_{ut}+F^{cc}_{tu} F^{cc}_{ut}\big)   \right\} \, ,
\label{qq}
\end{eqnarray}
whereas for different flavors $qq'\rightarrow qq'$
\begin{eqnarray}
|{\cal M}(qq'\rightarrow qq')|^2 = g^4 \,
{2\over 9}{1\over t^2} \, \Big[\big(sF^{bb}_{tu}\big)^2  +\big(sG^{cc'}_{tu}\big)^2 +\big(uG'^{bc}_{tu}\big)^2  +\big(uG'^{bc'}_{tu}\big)^2     \Big] ,
\label{qq'}
\end{eqnarray}
where
\begin{eqnarray}
F_{tu}^{bb} & = &\frac{t}{g^2}V_{abab}(-t/M_s^2,-u/M_s^2) \nonumber\\
G_{tu}^{bc} & = &\frac{t}{g^2}V_{abac}(-t/M_s^2,-u/M_s^2) \nonumber\\
G_{tu}^{'bc} & = &\frac{t}{g^2}V'_{abac}(-t/M_s^2,-u/M_s^2) \, ,\label{FGfun}
\end{eqnarray}
with the $V$-functions on the r.h.s.\  defined in Eq.(5.71) of
Ref. \cite{Lust:2008qc}.
Note that the above definitions single out the QCD coupling constant $g\equiv
g_a$, although the $V$-functions depend on the geometry of all internal cycles
associated D-branes. For example,
\begin{equation}
V_{abab}(t,u)=2\pi \a'g_s\int^1_0dx \;x^{t-1}(1-x)^{u-1}\ I(x)\sum
_{p_a,p_b\in {\mathbb{Z}} } e^{-S^{ba}_{\rm{inst.}}(x)},
\end{equation}
where $I(x)$ is the ``quantum'' part of the four--fermion string
 amplitude while the exponential
 $e^{-S^{ba}_{\rm{inst.}}(x)}$ weights the instanton contributions of D-branes
 wrapping various internal cycles, with $p_a$ and $p_b$ counting the
 winding numbers of $a$- and $b$-stacks along the respective cycles.
The function $V_{abab}(t,u)$
has kinematic poles in $t$- and  $u$-channels, originating from
$x\to 0$ and $x\to
1$ integration limits, respectively.  Near $x=0$, it is necessary to perform a
 Poisson
resummation of $p_a$ winding numbers. There are two consequences of this operation:
appearance of the  $(V_p^a)^{-1}$
factor which converts the string coupling $g_s$ into $g^2 \equiv g_a^2$ of
the $a$-stack (QCD), see Eq.(\ref{Dpgaugecoupling}), and exposition of
 $t$-channel
poles associated to the exchanges of massless gluons and of their Kaluza-Klein
excitations that can also wind around $\Sigma_b$ cycles of the other stack.
The masses of these particles are given
in Eq.~(5.50) of Ref.\cite{Lust:2008qc} and can be schematically written as:
\begin{equation}
M_{ba}^2= ( M_{KK}^{(a)})^2+(M_{\rm wind.}^{(b)} )^2\ .
\end{equation}
Note that in the process under consideration, these poles appear
outside the physical region of $t\leq 0$, however their effects may compete
with the string states if
$M_{ba}<M_s$.
Hence it is important to understand under what circumstances the
corresponding contributions to the scattering amplitudes can be neglected.
One possibility is to have the QCD Kaluza-Klein threshold above the string
scale, $V_p^a<(M_s)^{-p}$, with other cycles being larger, {\em i.e}. $V_p^b>(M_s)^{-p}$.
This is certainly consistent with $g_a>g_b$, {\em i.e.}, with the QCD coupling being the
strongest one. However, it is not possible to accomplish the suppression of {\em all\/}
Kaluza-Klein particles in this way, for the following reason.  Near $x=1$,
Poisson resummation in $p_b$ brings the volume factor $(V_p^b)^{-1}$
which converts the string coupling $g_s$ into $g_b^2$ and exposes  the (unphysical) $u$-
channel poles at the masses
\begin{equation}
M_{ab}^2= (M_{KK}^{(b)} )^2+( M_{\rm wind.}^{(a)} )^2\ ,
\end{equation}
associated to Kaluza-Klein excitations of electro-weak vector bosons
(possibly winding along the QCD cycles). Thus the same volume hierarchy that lifts
the QCD KK threshold be{\nolinebreak}yond the string scale brings down the
winding states of electro-weak gauge bosons below the
string scale. The ``consolation prize'' is that the corresponding
amplitudes come with the smaller electro-weak coupling $g_b^2$.
In general, any hierarchy will leave some KK or winding states below the string scale.

We find that quark-quark scattering amplitudes, which do not exhibit
resonant behav{\nolinebreak}ior in the physical domain of kinematic
variables (except for the massless $t$- and $u$-channel poles), are particularly
sensitive to a possible
hi{\nolinebreak}erarchy of various compactification cycles because KK
and winding modes can produce a considerable ``continuum'' background below the
string threshold. In order to obtain a rough numerical estimate of
such effects, we assume $M_{ab}<M_s<M_{ba}$ and take the $\alpha'\to
0$ limit of Eq.(\ref{FGfun}), keeping only the contribution of the
lightest excitations of electro-weak bosons:
\begin{equation}
F_{tu}^{bb} = 1+{g_b^2 t\over g_a^2 u}+  {g_b^2 t\over g_a^2 }\
{N_p\,\Delta\over u-M_{ab}^2}\ .
\label{F27}
\end{equation}
The factor $N_p$ takes into account a possible degeneracy of the lowest level while
$\Delta\sim e^{-M_{ab}^2/M_s^2}$ takes into account the effects of finite
thickness of the brane intersections.
Any realistic model must contain D-branes with at least three extra spatial
dimensions, therefore it is reasonable to set $N_p=3$.
D-brane thickness effects are also model-dependent,\footnote{In the intersecting
brane models discussed in Ref.\cite{Lust:2008qc}, $\Delta=\delta^{-M_{ab}^2/M_s^2}$, with
$\ln\delta=2\psi(1)-\psi(\theta/\pi)-\psi(1-\theta/\pi)$, where $\theta$ is the
angle between electro-weak and QCD stacks. Note that $\delta=16$ for orthogonal stacks.}
but they are exponentially suppressed in the limit $M_{ab}\ll M_s$. In our numerical
analysis, we will be considering $M_{ab}=0.7 M_s$, assumming that such KK scale is
sufficiently low for ignoring the effects of D-brane thickness and setting $\Delta=1$.
Futhermore, we set
\begin{equation}
G_{tu}^{bc}=
G_{tu}^{'bc}= 1\ .
\end{equation}

\subsection{String Theory {\em vs.} Field Theory $\bm{\mapsto \ U(N)}$ {\em vs.}
$\bm{SU(N)}$}

As discussed in the Introduction, the gauge group associated with the
color and electroweak stacks are $SU(3)_{\rm C} \times U(1)_a$ and
$SU(2)_{\rm EW} \times U(1)_b$. These extra $U(1)$'s are a source of
dramatic differences in the scattering amplitudes from the field
theoretic expectation. A striking example of this can be seen in the $qq'\to qq'$ process described by
Eqs.~(\ref{qq'}) and (\ref{F27}). The first term in (\ref{F27}), the
``1,'' describes $t$-channel gluon exchange of an $SU(3)$ gluon,
whereas the second term is the contribution from the $u$-channel exchange
of a charged $SU(2)$ $W^\pm$. What is apparently missing is the
$t$-channel exchange of the neutral $W^0$, which is certainly present
in the $SU(2)$ field theory.

The source of this effect can be traced to cancelations of the
$U(1)_b$ and $SU(2)$ contributions. In Ref.~\cite{Lust:2008qc}, a subtle
procedure was used for separating $SU(N)$ from $U(1),$ based on the
identity
\begin{equation}
\delta^{{\alpha}_1}_{\alpha_2}\delta^{\alpha_3}_{\alpha_4}
=2\sum_a
(T^{a})^{\alpha_1}_{\alpha_4}
(T^{a})^{\alpha_3}_{\alpha_2}+2\,Q^2_A\delta^{\alpha_1}_{\alpha_4}
\delta^{\alpha_3}_{\alpha_2} \ ,\label{groupf}\end{equation}
where the sum is over all $SU(N)$ generators and $Q_A=1/\sqrt{2N}$ is the properly normalized $U(1)$ charge.
This was done for the amplitude involving $SU(2)$ doublets $q^{\alpha}_{\beta}$ labeled by $U(2)$ incides $\beta$ and $U(3)$ (color) indices $\alpha$:
\begin{equation}
M(q^{\al_1}_{\be_1}q^{\al_2}_{\be_2}\to q^{\al_3}_{\be_3}q^{\al_4}_{\be_4})\sim R(t,u)\,
\delta^{\al_1\al_4}\delta^{\al_2\al_3}\times
\delta^{\be_1\be_3}\delta^{\be_2\be_4}+R(u,t)\,
\delta^{\al_1\al_3}\delta^{\al_2\al_4}\times
\delta^{\be_1\be_4}\delta^{\be_2\be_3} \,
\label{majorTom}
\end{equation}
where
\begin{equation}
R(t,u)= {g_a^2\over t} +{g_b^2\over u}+\dots
\end{equation}
Now consider
the process $u^{\al_1}d^{\al_2}\to u^{\al_3}d^{\al_4}$.
According to the string-theoretical Eq.(\ref{majorTom}), only the
first term contributes in this case. There is a gluon pole in the
$t$-channel and an electro-weak (charged $W$) pole in the $u$-channel,
captured by the $F_{tu}$ term in (\ref{qq'}). So what has happened to
the $W^0$ exchange in the $t$-channel?  {\em A priori}, it is contained in the second term of Eq.(\ref{majorTom}). Indeed, by using Eq.(\ref{groupf}), its electro-weak $U(2)_b$ group factor can be rewritten  as:
\begin{equation}
\delta^{\be_1\be_4}\delta^{\be_2\be_3}=2 (T^1)^{\be_1\be_3}(T^1)^{\be_2\be_4}+
2 (T^2)^{\be_1\be_3}(T^2)^{\be_2\be_4}+2 (T^3)^{\be_1\be_3}(T^3)^{\be_2\be_4}
+\frac{1}{2}I^{\be_1\be_3}I^{\be_2\be_4}\ ,
\label{deldel}\end{equation}
with the third term on the r.h.s.\ representing $W^0$ exchange between $u$ and $d$ quarks. Furthermore, the kinematic factor $R(u,t)$ supplies the corresponding $g_b^2/t$ pole term. However, as it is made explicit in Eq.(\ref{deldel}), this $W^0$ contribution is canceled  by the contribution of the extra $Z'$ gauge boson associated to $U(1)_b$.
Note that $U(1)_b$ is anomalous and, in the context of realistic models, it combines with other anomalous $U(1)$'s
to anomaly-free hypercharge $U(1)_Y$ and a number of other $U(1)$'s
coupled to anomalous currents. At the end of the day, the corresponding $Z'$-bosons become massive~\cite{Ghilencea:2002da}. This suggests
removing``by hand'' the last term from the r.h.s.\ of Eq.(\ref{deldel}), which leaves $W^0$ as the only (massless) neutral particle exchanged in the $t$-channel. (Because of the flavor assignments for $ud\rightarrow ud$, the first two terms in Eq.(\ref{deldel}) are zero.) However,
such an {\em ad hoc} procedure introduces its own complications in
the form of a QCD-strength ($\sim g_a^2/u$) pole in the $u$-channel, because this unwanted term cannot be separated from the desired $t$-channel pole if one wants to preserve the $t$-$u$ channel string duality of $R(u,t)$. In the
following section we will address phenomenological approaches to
extract information on these interesting aspects of four-fermion
amplitudes.


\section{Dijet Angular Distributions}

QCD parton-parton cross sections are dominated by $t$-channel
exchanges that produce dijet angular distributions which peak at small
center of mass scattering angles, $\theta^*$. In contrast,
non--standard contact interactions or excitations of resonances result
in a more isotropic distribution. In terms of rapidity variable for
standard transverse momentum cuts, dijets resulting from QCD processes
will preferentially populate the large rapidity region, while the new
processes generate events more uniformly distributed in the entire
rapidity region. To analyze the details of the rapidity space the D\O\
Collaboration introduced a new parameter~\cite{Abbott:1998wh},
\begin{equation}
R = \frac{d\sigma/dM|_ {(|y_1|,|y_2|< 0.5)}}{d\sigma/dM|_{(0.5 < |y_1|,|y_2| < 1.0)}} \, ,
\end{equation}
the ratio of the number of events, in a given dijet mass bin, for both
rapidities $|y_1|, |y_2| < 0.5$ and both rapidities $0.5 < |y_1|,
|y_2| < 1.0$. Here, $M$ is the dijet invariant mass.  With the
definitions $Y\equiv \thalf (y_1 + y_2)$ and $y \equiv
\thalf(y_1-y_2)$, the cross section per interval of $M$ for $p
p\rightarrow {\rm dijet}$ is given by
\begin{eqnarray}
\frac{d\sigma}{dM} & = & M\tau\ \sum_{ijkl}\left[
\int_{-Y_{\rm max}}^{0} dY \ f_i (x_a,\, M)  \right. \ f_j (x_b, \,M ) \
\int_{-(y_{\rm max} + Y)}^{y_{\rm max} + Y} dy
\left. \frac{d\sigma}{d\hat t}\right|_{ij\rightarrow kl}\ \frac{1}{\cosh^2
y} \nonumber \\
& + &\int_{0}^{Y_{\rm max}} dY \ f_i (x_a, \, M) \
f_j (x_b, M) \ \int_{-(y_{\rm max} - Y)}^{y_{\rm max} - Y} dy
\left. \left. \frac{d\sigma}{d\hat t}\right|_{ij\rightarrow kl}\
\frac{1}{\cosh^2 y} \right]
\label{longBH}
\end{eqnarray}
where $\tau = M^2/s$, $x_a =
\sqrt{\tau} e^{Y}$,  $x_b = \sqrt{\tau} e^{-Y},$
and
\begin{equation}
  |{\cal M}(ij \to kl) |^2 = 16 \pi \hat s^2 \,
  \left. \frac{d\sigma}{d\hat t} \right|_{ij \to kl} \, .
\end{equation}
In this section we reinstate the caret notation ($\hat s,\ \hat t,\
\hat u$) to specify partonic processes. The $Y$ integration range in
Eq.~(\ref{longBH}), $Y_{\rm max} = {\rm min} \{ \ln(1/\sqrt{\tau}),\ \
y_{\rm max}\}$, comes from requiring $x_a, \, x_b < 1$ together with
the rapidity cuts $y_{\rm min} <|y_1|, \, |y_2| < y_{\rm max}$. The
kinematics of the scattering also provides the relation $M = 2p_T
\cosh y$, which when combined with $p_T = M/2 \ \sin \theta^* = M/2
\sqrt{1-\cos^2 \theta^*},$ yields $\cosh y = (1 - \cos^2
\theta^*)^{-1/2}$.  Finally, the Mandelstam invariants occurring in
the cross section are given by $\hat s = M^2,$ $\hat t = -\thalf M^2\
e^{-y}/ \cosh y,$ and $\hat u = -\thalf M^2\ e^{+y}/ \cosh y.$ The
ratio $R$ is a genuine measure of the most sensitive part of the
angular distribution, providing a single number that can be measure as
a function of the dijet invariant mass~\cite{Meade:2007sz}.

In assessing the four-fermion amplitudes there are two independent
contributions of stringy physics: {\em (i)} the presence of KK and
winding modes; {\em (ii)} anomalous effects in $qq' \to qq'$ because
of cancelations between $U(1)$ and $SU(N)$ contributions. The latter
acquires complexity because the unknown anomalous masses of the $U(1)$
gauge bosons disturb this cancelation. In order to simplify the
analysis we present two extreme cases.  Firstly, we include a single
KK contribution in $qq \to qq$, while omitting the non-QCD part of $q
q' \to q q'$ which contains the anomalous $U(1)$ cancelation. The
importance of this omission is somewhat mitigated because the $qq'$
valence quark luminosity is only $4/5$ of the $qq$ luminosity. Secondly, we
go to the opposite extreme case by omitting the KK contribution and
including the $q q' \to qq'$ amplitude in the limit where all the
gauge bosons are massless.

\begin{figure}[tbp]
\begin{minipage}[t]{0.49\textwidth}
\postscript{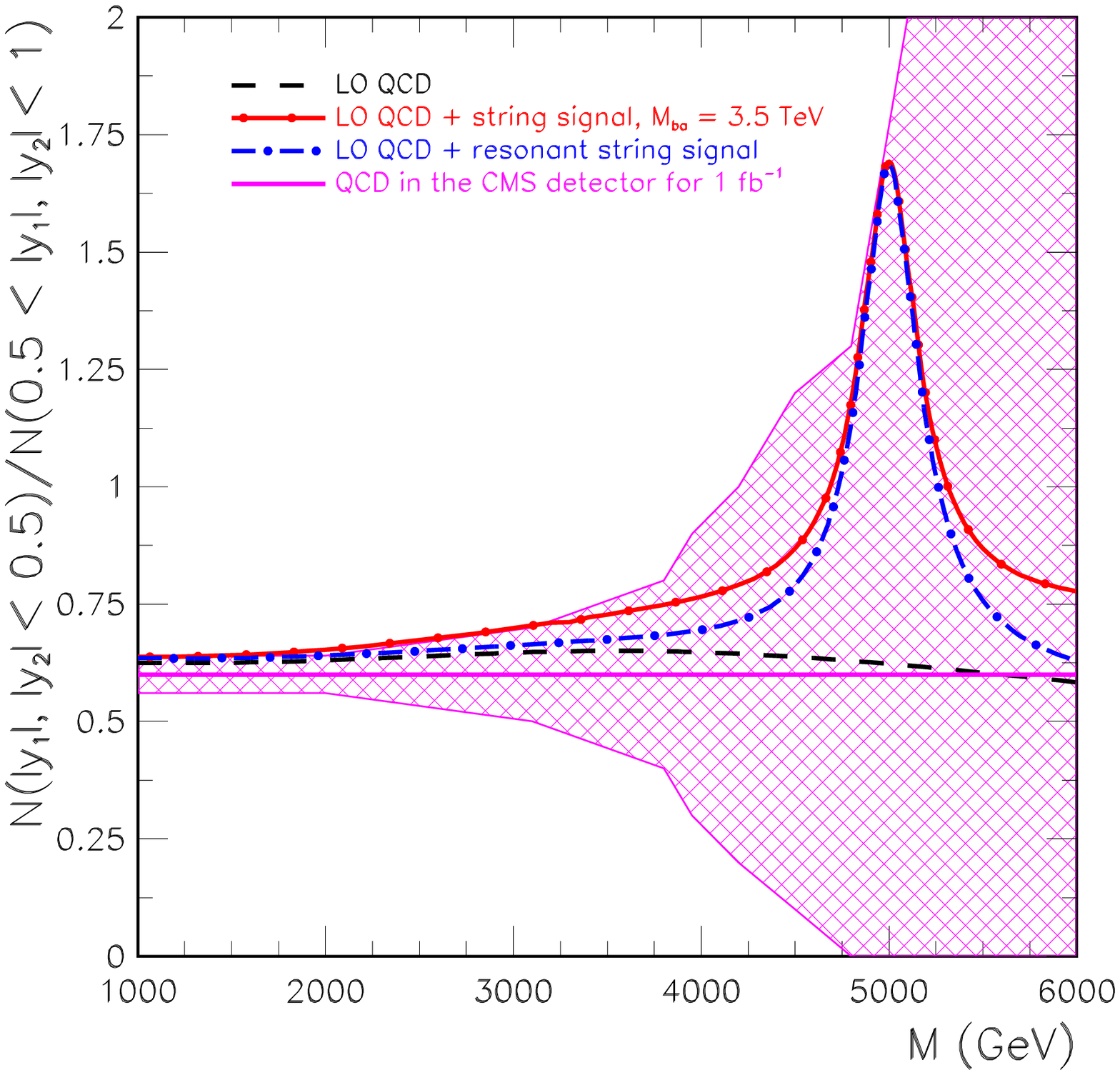}{0.99}
\end{minipage}
\hfill
\begin{minipage}[t]{0.49\textwidth}
\postscript{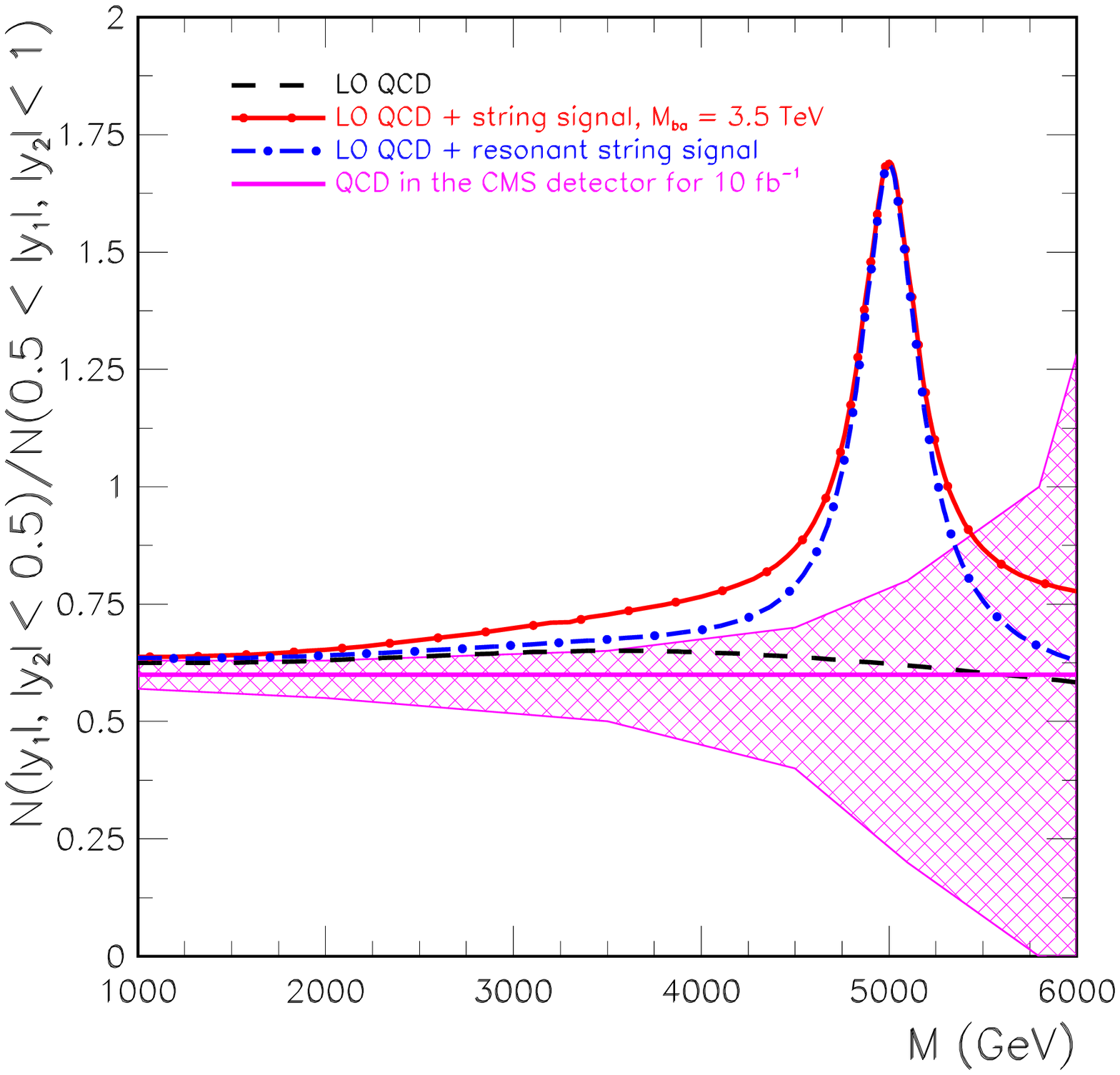}{0.99}
\end{minipage}
\caption{For integrated luminosities of 1~fb$^{-1}$ (left panel) and
  10~fb$^{-1}$ (right panel) the expected value (solid line) and
  statistical error (shaded region) of the dijet ratio of QCD in the
  CMS detector~\cite{Esen} is compared with LO QCD (dashed line), LO
  QCD + lowest massive string excitations (dot-dashed), and LO QCD +
  lowest massive string excitations + non-resonant string scattering
  in $qq \to qq$ (dot-solid). We have taken $M_{ab} = 3.5~{\rm TeV}$,
  $M_s = 5$~TeV, and $g^2_b/g_a^2 = 1/3$.}
\label{fig:1}
\end{figure}

\begin{figure}[tbp]
\begin{minipage}[t]{0.49\textwidth}
\postscript{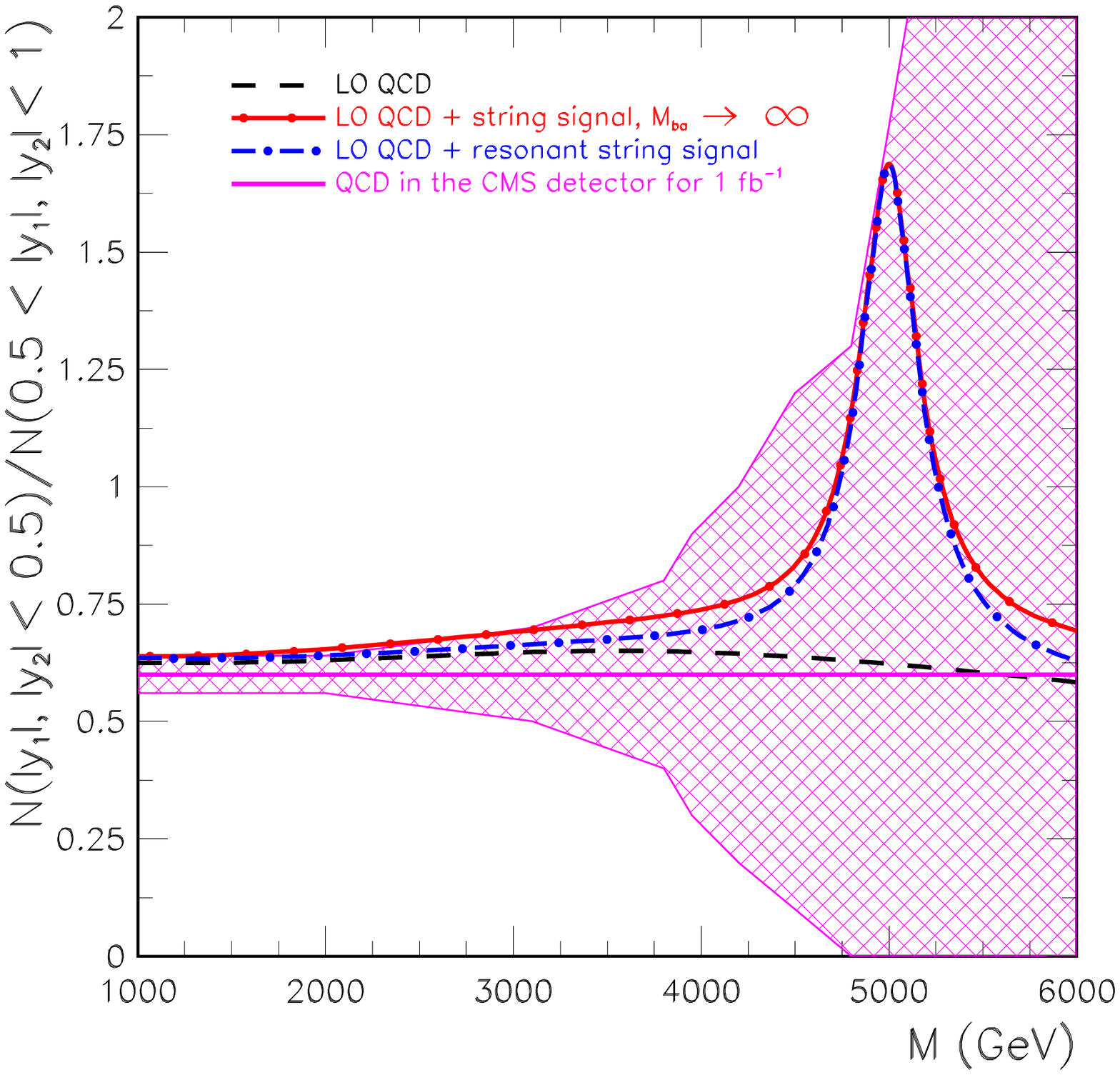}{0.99}
\end{minipage}
\hfill
\begin{minipage}[t]{0.49\textwidth}
\postscript{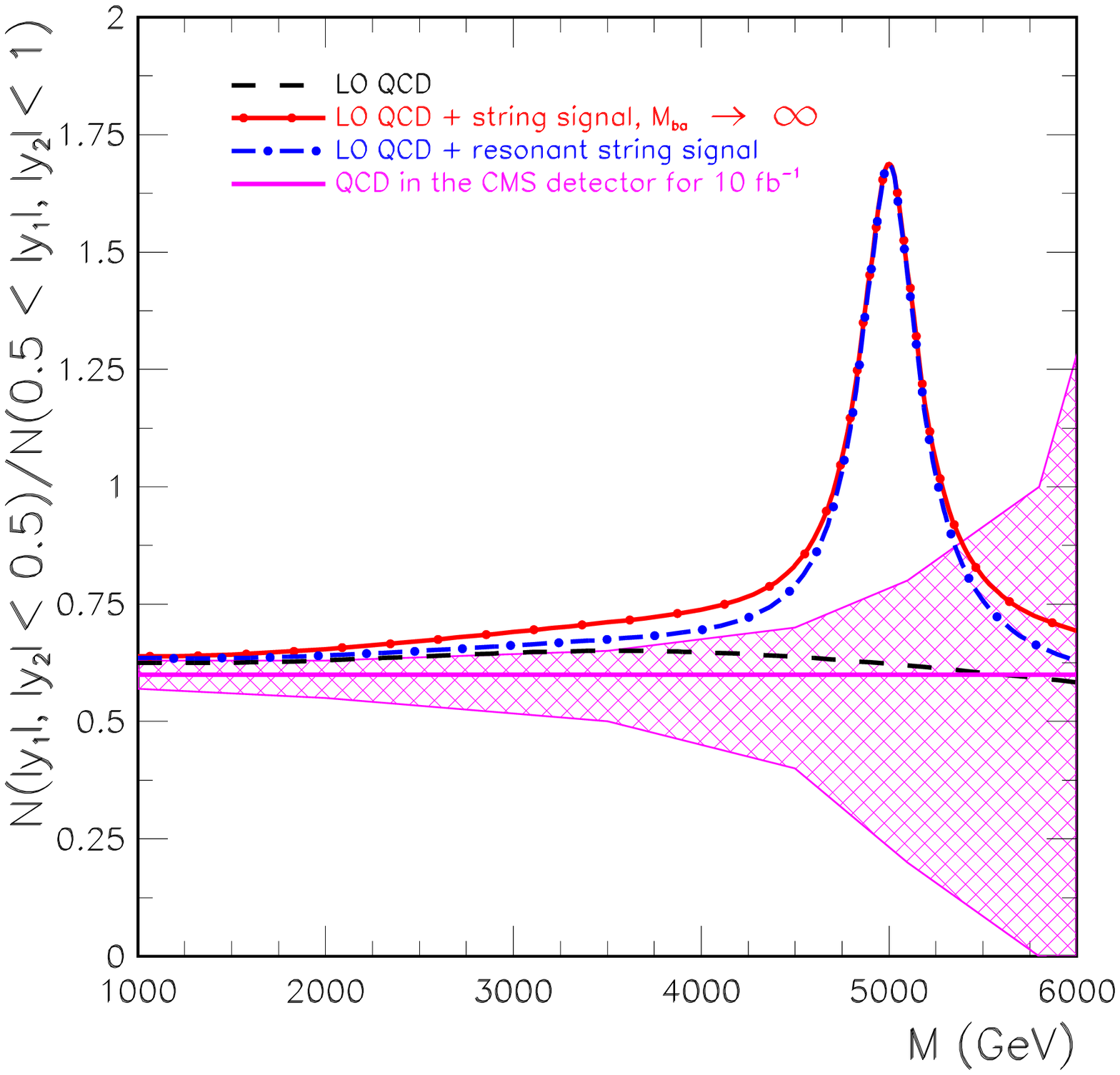}{0.99}
\end{minipage}
\caption{For integrated luminosities of 1~fb$^{-1}$ (left panel) and
  10~fb$^{-1}$ (right panel) the expected value (solid line) and
  statistical error (shaded region) of the dijet ratio of QCD in the
  CMS detector~\cite{Esen} is compared with LO QCD (dashed line), LO
  QCD + lowest massive string excitations (dot-dashed), and LO QCD +
  lowest massive string excitations + non-resonant string scattering
  in $qq \to qq$ and $qq' \to qq'$ (dot-solid), with all anomalous
  $U(1)$ masses set to zero (see text). We have taken $M_{ab} =
  \infty~{\rm TeV}$, $M_s = 5$~TeV, and $g^2_b/g_a^2 = 1/3$.}
\label{fig:2}
\end{figure}

In Figs.~\ref{fig:1} and \ref{fig:2} we compare the results from a
full CMS detector simulation of the ratio $R$, with predictions from
LO QCD, model-independent contributions to the $q^*$, $g^*$ and $C^*$
excitations, and the non-resonant string scattering. The synthetic
population was generated with Pythia, passed through the full CMS
detector simulation and reconstructed with the ORCA reconstruction
package~\cite{Esen}. For an integrated luminosity of 10~fb$^{-1}$ the
LO QCD contributions with $\alpha_{\rm QCD} = 0.1$ (corresponding to
running scale $\mu \approx M_s$) are within statistical fluctuations
of the full CMS detector simulation.  Since one of the purposes of
utilizing NLO calculations is to fix the choice of the running
coupling, we take this agreement as rationale to omit loops in QCD and
in string theory. It is clear from Fig.~\ref{fig:1} that incorporating
NLO calculation of the background and the signal would not
significantly change the deviation of the string contribution from the
QCD background.

In Fig.~\ref{fig:1} we show the contribution of the lightest KK mode (in the isotropic case with lowest KK level degeneration $N_p=3$) for $M_{ab} = 3.5$~TeV. In Fig.~2 we show the anomalous effect in $qq' \to qq'$ amplitude in the limit where all the gauge bosons are massless. We note that inclusion of anomalous masses reinstitute the $W^0$ pole in the $qq'$ amplitude and this has the effect of decreasing the deviation of $R$ from its QCD behavior. In that case we expect the deviation to be dominated by the finite mass KK contribution, shown in Fig.~\ref{fig:1}. In order to assess the statistical significance of this claim, we calculate the signal-to-noise ($S/\sqrt{B}$) ratio for events with both rapidities between (0, 0.5) and invariant mass in the interval $3~{\rm TeV} < M < 3.5~{\rm TeV}.$ For $10$~fb$^{-1}$ we find the following contributions with respect to the SM background ($B$): {\it (i)} from the tail of the Regge excitation $S/\sqrt{B} = 100/48 = 2 \sigma$; {\it (ii)} from the KK modes $S/\sqrt{B} = 290/48 = 6 \sigma$. It is important to emphasize that the KK contribution viewed as a contact term is prescribed by the theory, in the form, the sign, and strength relative to the QCD and Regge contributions.

\section{Monojet, trijet, and jet plus dilepton configurations}

Events with a single jet plus missing energy ($\MET$) with balancing
transverse momenta (so-called ``monojets'') are incisive probes of new
physics, and therefore enormous effort has been dedicated to identify
the underlying model which can produce them~\cite{Rizzo:2008fp}. In
order to optimize our chances of uncovering low mass string
excitations, in what follows we analyze the predictions of D-brane
compactifications in the monojet channel.

As in the SM, the source of this topology is $ij \to k Z^0$ followed
by $Z^0 \to \nu \bar \nu.$ Both in the SM and string theory the cross
section for this process is of order $g^4$. Virtual KK graviton
emission ($ij \to k G$) involves emission of closed strings, resulting
in an additional suppression of order $g^2$ compared to $Z^0$
emission. A careful discussion of this suppression is given in
Ref.~\cite{Cullen:2000ef}. However, in some scenarios compensation for
this suppression can arise from the large multiplicity of graviton
emission, which is somewhat dependent on the cutoff
mechanism~\cite{Bando:1999di}. Additional evidence for string
excitations can be obtained through analysis of the related process
$Z^0 \to l^+ l^-.$ For $l = e,\ \mu$ the final topology is jet +
dilepton. For $l = \tau$, about two-thirds of the subsequent tau
decays are hadronic and so in 45\% of cases the observed configuration
is a trijet.

As in the dijet analysis we consider scattering processes on the (color) $U(3)_a$ stack of D-branes, leading now to $Z^0 +$ jet final state. Recall that the $U(1)_Y$ boson $Y_\mu$, which gauges the usual electroweak hypercharge symmetry, is a linear combination of $C_\mu$, the $U(1)_c$ boson $B_\mu$ terminating on a separate brane, and perhaps a third additional $U(1)_b$ (say $W_\mu$) sharing a $U(2)_b$ brane which is also a terminus for the $SU(2)_L$ electroweak gauge bosons $W_\mu^a.$ Thus, critically for our purposes, the $Z_\mu$ boson, which is a linear combination of $Y_\mu$ and $W^3_\mu$ will participate with the gluon octet in (string) tree level scattering processes on the color brane, processes which in the SM occur only at one-loop level.  Such a mixing between hypercharge and baryon number is a generic property of D-brane quivers, see {\it
  e.g}.\ Refs.~\cite{ant,bo,Berenstein:2006pk}.

By duplicating the discussion in Ref.~\cite{Anchordoqui:2007da} we obtain the
$s$-channel pole terms of the average square amplitudes contributing to
$pp \to Z^0$ + jet. The dominant terms are
\begin{eqnarray}
|{\cal M} (gg \to gZ)|^2  =  \frac{5}{3} \, Q^2 \,
\frac{g^4}{M_s^4}  \, \left[\frac{M_s^8}{(  s-M_s^2)^2
+ (\Gamma_{g^*}^{J=0}\ M_s)^2}  + \frac{  t^4+   u^4}{(  s-M_s^2)^2 + (\Gamma_{g^*}^{J=2}\ M_s)^2}\right]
\label{gggz}
\end{eqnarray}
and
\begin{eqnarray}
|{\cal M}(qg \to qZ)|^2   =   -\frac{1}{3} Q^2 \frac{g^4}{M_s^2}\
\left[ \frac{M_s^4   u}{(  s-M_s^2)^2 + (\Gamma_{q^*}^{J=\frac{1}{2}}\ M_s)^2}  \right.  + \left. \frac{u^3}{(s-M_s^2)^2 + (\Gamma_{q^*}^{J=\frac{3}{2}}\ M_s)^2}\right]
\label{qgqz}
\end{eqnarray}
where
\begin{equation}
Q = \sqrt{\tfrac{1}{6}} \ \kappa \ \sin \theta_W \approx 2.76\times
10^{-2}\ \left(\tfrac{\kappa}{0.14} \right)
\end{equation}
is the product of the $U(1)$ charge of the fundamental representation
($\sqrt{1/6}$) followed by successive projections onto the hypercharge
($\kappa$) and then onto the $Z^0$-boson ($\sin \theta_W$). In a self
evident notation the $\Gamma^J_{(q^*,g^*)}$'s are the total widths of the
participating resonances. These are calculated in
Ref.~\cite{Anchordoqui:2008hi}.

By convoluting with parton distribution functions we calculate
$d\sigma/dM$, where $M$ is the invariant mass of $Z^0 + $
jet~\cite{Roy:2009pc}. We then multiply by the corresponding branching
ratio ($Z^0 \to \nu\bar \nu$, $Z^0 \to e^+e^- + \mu^+\mu^-$, $Z^0 \to
\tau^+ \tau^-$) and integrate over the region $M_s - 2
\Gamma^J_{(q^*,g^*)} < M < M_s + 2 \Gamma^J_{(q^*,g^*)}$ (with the
appropriate width for each partonic subprocess) to obtain the resonant
cross section for the different topologies. For an integrated
luminosity of 100~fb$^{-1}$ one arrives at the signal-to-noise ratio
given in Table~I. The noise indicates the square root of the QCD
background number of events in each of these channels (for obtaining
the background in the resonant region we integrate over the interval
$M_s - 2 \Gamma_{\rm max} < M < M_s + 2 \Gamma_{\rm max}$, where
$\Gamma_{\rm max}$ is the largest width).

{}From Table~I we see that the LHC reach for identifying stringy
contributions from $pp \to Z^0 + $ jet is about $M_s = 3.5$~TeV, based
on monojets and jet + dilepton events. We can compare this with the
sensitivity previously obtained for $pp \to$ direct $\gamma$ + jet
($M_s = 5$~TeV) and in the dijet channel ($M_s = 6.8$~TeV, with $S/\sqrt{B} =
210/42$)~\cite{Anchordoqui:2008di}. Thus, if $M_s \alt 3.5$~TeV, the
$Z^0$ channel can serve as significant corroborative evidence for
stringy contributions, with the dijet channel clearly best suited for
discovery.

\begin{table}
  \caption{Signal-to-noise ratio ($S/\sqrt{B}$) of the jet + $\MET$, jet + dilepton, and
  trijet channels for an integrated luminosity of 100~fb$^{-1}$.}
\begin{tabular}{c|c|c|c}
\hline
\hline
~~~~$M_s$~(TeV)~~~~ & ~~~~jet + $\MET$~~~~ & ~~~~jet + dilepton~~~~ & ~~~~trijet~~~~ \\
\hline
1 & 21181/135  & 6989/78 & 1553/37 \\
2 & 719/31 & 273/18 & 53/9 \\
3 & 69/11 & 23/6 & 5/3 \\
4 & 9/4 & 3/2 & 1/1 \\
\hline
\hline
\end{tabular}
\end{table}

\section{LHC sensitivity for early discovery of string
  resonances}

The LHC is expected soon to begin circulation of beams for production
of $pp$ collisions at $\sqrt{s} = 10~{\rm TeV}$, followed by a physics
run at $\sqrt{s} = 14~{\rm TeV}$. In the LHC ramping up process it
will be crucial to observe various familiar SM processes
at their expected rates, with distributions and mass peaks at
previously measured values. Conventional wisdom holds that once
confidence in the ATLAS and CMS detectors has been deep-rooted,
the search for beyond SM physics will start. Of course,
new exotic phenomena could have cross sections so large, and
topologies so striking, that even a limited amount of collected data
and a non-ultimate detector performance could lead to exciting
results. In the spirit of~\cite{Gianotti:2005fm}, in this section we
explore the possibility of searching for lowest massive Regge
excitations of open strings {\em in parallel} with the calibration
phase. We will show that even with relatively poor knowledge of
detector response, searches of Regge recurrences could still be
possible.

\begin{figure}[tbp]
\postscript{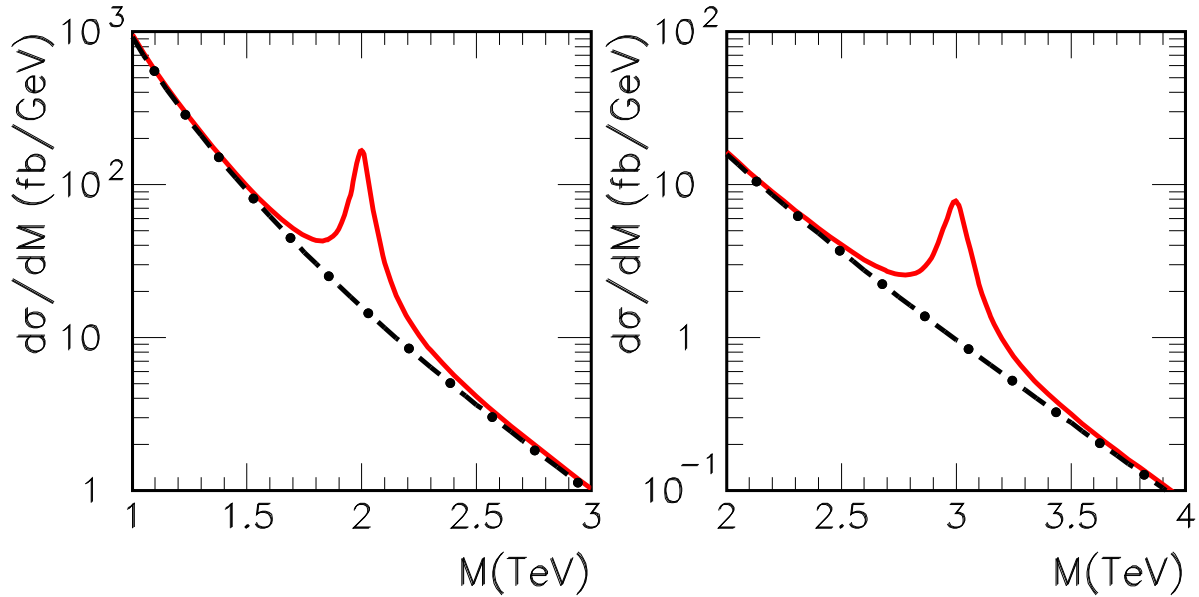}{0.99}
\caption{$d\sigma/dM$ (units of fb/GeV) {\em vs.}  $M$ (TeV) is
  plotted for the case of SM QCD background (dot-dashed line) and (first
  resonance) string signal + background (solid line), for $M_s = 2$~TeV
  (left) and $M_s = 3$~TeV (right).}
\label{fig:3}
\end{figure}

Following our previous analysis~\cite{Anchordoqui:2008di}, we study
the discovery reach of Regge excitations in data binned according to
the invariant mass $M$ of the dijet, after setting cuts on the
different jet rapidities, $|y_1|, \, |y_2| \le 1$~\cite{CMS} and
transverse momenta $p_{\rm T}^{1,2}>50$ GeV.  In Fig.~\ref{fig:3} we
show representative plots of the invariant mass spectrum, for $M_s
=2$~TeV and $M_s = 3$~TeV. The string signal has been calculated using
(\ref{longBH}) with the corresponding $2 \to 2$ scattering amplitudes
given in Eqs. (\ref{gggg2}) - (\ref{qgqg2}). The QCD background has
been calculated at the partonic level from the same processes as
designated for the signal, with the addition of $qq\to qq$ and $q \bar
q \to q \bar q$.

We now estimate (at the parton level) the early LHC discovery
reach. Standard bump-hunting methods, such as obtaining cumulative
cross sections, $\sigma (M_0) = \int_{M_0}^\infty \frac{d\sigma}{dM}
\, \, dM$, from the data and searching for regions with significant
deviations from the QCD background, may reveal an interval of $M$
suspected of containing a bump.  With the establishment of such a
region, one may calculate a signal-to-noise ratio with the signal rate
once more estimated in the invariant mass window $[M_s - 2 \Gamma, \,
M_s + 2 \Gamma]$, and $\sqrt{B}$ taken in the same dijet mass interval
for the same integrated luminosity.

For $M_s = 3$~TeV and 100~pb$^{-1}$ of data collected at $\sqrt{s} =
10$~TeV, we expect $S/\sqrt{B} = 127/20 = 6
\sigma.$ For an overly conservative assumption of integrated
luminosity $\approx 10~{\rm pb}^{-1}$, a $S/\sqrt{B} = 204/19 >
10\sigma$ is expected for string scales as high as $M_s = 2$~TeV. It
is remarkable that within 1 year of data collection at $\sqrt{s} =
10$~TeV, {\it string scales as large as 3~TeV are open to discovery at
  the $\geq 5\sigma$ level.}  Once more, we stress that these results
contain no unknown parameters. They depend only on the D-brane
construct for the SM, and {\it are independent of
  compactification details.}

\section{Conclusions}

\begin{itemize}

\item In the framework of D-branes, we have reviewed the theoretical
  complexity of four-fermion scattering amplitudes in low scale string
  theory, focussing on the aspects which are central to a
  phenomenological analysis. These consist of a significant interplay
  among string Regge recurrences, Kaluza-Klein modes, and D-brane
  winding modes. The relative contributions of KK and winding modes
  also reflect the irreducible correlation between strong and
  electroweak interactions inherent in the intersecting brane picture.

\item Although there are no $s$-channel resonances in $qq\rightarrow
  qq$ and $qq'\rightarrow qq'$ scattering, KK modes in the $t$ and $u$
  channels generate calculable effective 4-fermion contact
  terms. These in turn are manifest in an enhancement in the
  continuum below the string scale of the
  $R$ ratio for dijet events, described in the text. For $M_{\rm
    KK}\le 3$ TeV, this contribution can be detected at the LHC with 6$\sigma$
  significance above SM background.  In combination with the
  simultaneous observation in dijet events of a string resonance at
  $M_s> M_{\rm KK}$, this would consolidate the stringy interpretation
  of these anomalies.

\item Analyses of the various final-state topologies arising in the
  $pp \to Z^0+$ jet channel show that, for $M_s <3.5$~TeV, monojet,
  trijet, and dilepton plus jet configurations could provide
  corroborative evidence for Regge excitations in D-brane models.

\item Excitation cross sections for Regge recurrences in the early LHC
  run, operating at 5~TeV per beam are sizeable. Consequently, dijet
  signals of string excitations may be so large for $M_s \alt 3~{\rm
    TeV}$ that they could be detected at statistically significant
  event rates in the first year of running.

\end{itemize}

\section*{Acknowledgments}

We are grateful to Lance Dixon for continuous inspiring discussions.
L.A.A.\ is supported by the U.S. National Science Foundation Grant No
PHY-0757598, and the UWM Research Growth Initiative.  H.G.\ is
supported by the U.S. National Science Foundation Grant No
PHY-0757959.  The research of D.L.\ and St.St.\ is supported in part
by the European Commission under Project MRTN-CT-2004-005104.  S.N. is
supported by the UWM Research Growth Initiative.  He is grateful to
Max--Planck--Institut f\"ur Physik in M\"unchen, for their kind
hospitality. The research of T.R.T.\ is supported by the U.S.
National Science Foundation Grants PHY-0600304, PHY-0757959 and by the
Cluster of Excellence ``Origin and Structure of the Universe'' in
M\"unchen, Germany. Any opinions, findings, and conclusions or
recommendations expressed in this material are those of the authors
and do not necessarily reflect the views of the National Science
Foundation.

\end{document}